\documentclass[traditabstract]{aa}  
\usepackage{graphicx}
\usepackage{txfonts}
\usepackage[pdftex=true,colorlinks=true,linkcolor=blue,citecolor=blue,pdfborder=
            {0 0 0},bookmarks=false,urlcolor=blue,pdftitle={Modeling and
            Analysis Generic Interface for eXternal numerical codes (MAGIX)},
            pdfsubject={},pdfauthor={},pdfkeywords={}]{hyperref}

\usepackage[numbers,curly,sort]{natbib}
\bibpunct{(}{)}{;}{a}{}{,}                             
\begin{document}
    \title{Modeling and Analysis Generic Interface for eXternal numerical codes
           (MAGIX)\thanks{http://www.astro.uni-koeln.de/projects/schilke/MAGIX}}


    \author{T.~M\"{o}ller\inst{1}
            \and
            I.~Bernst\inst{1}
            \and
            D.~Panoglou\inst{1}
            \and
            D.~Muders\inst{2}
            \and
            V.~Ossenkopf\inst{1}
            \and
            M.~R\"{o}llig\inst{1}
            \and
            P.~Schilke\inst{1}
            }

    \institute{I. Physikalisches Institut, Universit\"{a}t zu K\"{o}ln,
               Z\"{u}lpicher Str. 77, D-50937 K\"{o}ln, Germany\\
               \email{moeller@ph1.uni-koeln.de}
               \and
               Max-Planck-Institut f\"{u}r Radioastronomie, Auf dem H\"{u}gel
               69, 53121 Bonn, Germany\\
              }

    \date{Received August 25, 2012 / Accepted October 13, 2012}

    \abstract
{The modeling and analysis generic interface for external numerical codes
(MAGIX) is a model optimizer developed under the framework of the coherent set
of astrophysical tools for spectroscopy (CATS) project. The MAGIX package
provides a framework of an easy interface between existing codes and an
iterating engine that attempts to minimize deviations of the model results from
available observational data, constraining the values of the model parameters
and providing corresponding error estimates. Many models (and, in principle,
not only astrophysical models) can be plugged into MAGIX to explore their
parameter space and find the set of parameter values that best fits
observational/experimental data. MAGIX complies with the data structures and
reduction tools of ALMA (Atacama Large Millimeter Array), but can be used with
other astronomical and with non-astronomical data.}


    \titlerunning{MAGIX}
    \authorrunning{T.~M\"{o}ller \textit{et al.}}

    \maketitle

    \section{Introduction}

Most physical or chemical models use a set of parameters. Finding the best
description of observational/experimental data using a certain model implies
determining the parameter set that most closely reproduces the data by some
criteria, typically the minimum of a merit function. Often the $\chi^2$
distribution\footnote{The $\chi^2$ distribution is a function of relative
quadratic differences between experimental and model values.} is used, and we
will use this term throughout, although it should be understood
that it could be replaced by other appropriate merit functions. Other important
results are the goodness of fit, in absolute terms, and confidence levels for
determined parameters. This is a generic problem independent of the actual
model, and instead of implementing an optimizer in each and every program,
parameter optimization can be separated. Therefore, a software package is needed
that finds the best parameter set in an iterative procedure for arbitrary
models by comparing the results of the physical model for a given parameter
set with the experimental data set and modifying the parameter set to improve
the quality of the description, i.e., by reducing the value of $\chi^2$. In
general, the physical and chemical models are multidimensional non-linear
functions of the input parameters. Thus, finding the best description for a
given experimental data set means finding the global minimum of the $\chi^2$
function, which is a function of the model input parameters.

Many optimization functions will find minima, but they could be local minima,
which do not describe the results adequately, and can lead to misleading
interpretations. Therefore, one has to find the global minimum of the $\chi^2$
function to obtain a good description of the experimental data. However,
finding the global minimum of an arbitrary function is challenging and
has been practically impossible for many problems so far. To circumvent this, we
need algorithms that allow us to explore the landscape of the $\chi^2$ function
and calculate probabilities for the occurrence of minima. Combining certain
algorithms and making use of the different advantages of the applied algorithms
allows a reliable but not absolutely unique interpretation of the experimental
data. Most of the algorithms are very computationally expensive, and the
computational effort tends to scale with the degree of reliability.

These requirements are very general. Hence it makes sense to generate a package
that is able to read in experimental data, communicate with any registered
external model program\footnote{Here, the phrase ``external model program''
means the external program that calculates the model function depending on
several input parameters.} and compare automatically the result of the physical
model with the given data through the figure of merit. It should improve the
quality of the fit within an iterative procedure by adjusting the input
parameters using several algorithms that fulfill the wide range of requirements
mentioned above.

To make MAGIX as flexible as possible, we developed it as a stand-alone program
instead of a library for a certain programming language. A library is always
coupled to a certain language and requires knowledge of their usage. A user
without sufficient experience would not benifit from MAGIX while a stand-alone
program requires only the knowledge of how to start the model program. No
further experience in software programming is necessary. Additionally, many
model programs such as Radmc-3D or Lime are controlled by input files or by
partial modification of their source code. Therefore, writing one or more
files on the disk, starting the model program, and finally reading in the result
is inevitable. In addition, MAGIX offers the possibility to use a so-called RAM
disk (or RAM drive), which is orders of magnitude faster than a normal hard
disk. By using an RAM disk, the function evaluation / model computation becomes
the dominant part in the whole process for nearly all external model programs.
Therefore our approach will likely be more beneficial for the majority of
users.

In the following we describe the structure and functionality of the MAGIX
package that implements this system. We start with an overview of the different
parts of MAGIX, followed by a detailed description of the so-called registration
process that couples the extended model to the fitter code. In
\S\ref{sec:optimization} we explain the different algorithms included in the
MAGIX package in more detail and end with an astrophysical application of MAGIX
to absorption lines toward SgrB2(M) from the HEXOS~GT~KP \citet{Schilke1}
\footnote{An earlier version of this code was written by \citet{Dalia}.}.

    \section{MAGIX}

MAGIX provides a generic toolbox for modeling data through external model
programs. This requires registering the model, i.e., telling MAGIX about
the input and output formats, and how to run the model. This has to be done only
once for each model. Then, for each run, the model has to be instantiated,
i.e., given the initial values and parameter boundaries. After an iterative
procedure MAGIX provides the best-fit parameters, within the framework of the
model, to a particular data set, including confidence intervals for the
parameters. Internally, the parameter and model sets are stored in XML files,
but this is transparent to the user. MAGIX consists of a frontend to create
and manipulate the necessary XML files, modules for reading experimental data,
the fitting engine, and an output module.

All required XML files can be created from a GUI, which aspires to be
self-explicable. The various XML files are:
\begin{itemize}
  \item The registration file, which contains a description of the structure of
        the input and output file(s) of the external model program (see
        \S\ref{sec:registration}).
  \item The so-called instance file, which includes the names, initial values
        (and ranges) for all model parameters for a particular fit. It works in
        conjunction with the registration file and indicates the model
        parameters that should be optimized by MAGIX and those that are to be
        kept fixed.
  \item The XML file containing settings for the import/export of experimental
        data, i.e., path(s) and name(s) of the data file(s), format(s),
        range(s), etc.
  \item The so-called algorithm control file, that defines which algorithm or
        algorithmic sequence MAGIX should use for the optimization together
        with the corresponding settings for the different algorithms (see
        \S\ref{sec:optimization}).
  \item The I/O control file, including the paths and file names of the
        aforementioned XML files.
\end{itemize}
The final output of MAGIX includes the best-fit model, the best-fit
parameters, and an estimate of the goodness of fit. Depending on the method
chosen, it provides an exploration of the parameter space, i.e., information
about the existence of other minima, and, optionally, the confidence intervals.

Depending on the model and the chosen algorithm, the computational load is heavy
(sometimes hundreds or thousands of function calls are necessary).
\texttt{OpenMP}\footnote{\url{http://openmp.org/wp/}} parallelization for
simultaneous evaluations of the external program is available for most of the
algorithms (except for interval-nested sampling) if the external model program
fulfills certain requirements described in \S\ref{sec:registration}.

The \texttt{OpenMP} parallelization is used whenever the $\chi^2$ values
for a set of independent so-called parameter vectors have to be calculated. (A
parameter vector contains a value for each free parameter and describes a point
within the parameter space). Within such a parallelized loop all subroutines
used to calculated the $\chi^2$ value of a certain parameter vector such as
writing the input file(s), reading in the output file(s) and calculating the
corresponding $\chi^2$ value are executed simultaneously. The system overhead
is negligible for external model programs that require more than a few
milliseconds for one function evaluation. Furthermore, the usage of an RAM disk
as described above could reduce the system overhead as well.

Additionally, MAGIX creates three types of log-files for each algorithm call:
The first log-file contains the lowest $\chi^2$ value with the corresponding
values of the free parameters for each iteration step. The second log-file
contains the corresponding input file(s) for the external model program
for each iteration step. The comprehensive third log-file contains the $\chi^2$
values and their corresponding free parameter values of all function calls.

A detailed description of all parts of MAGIX can be found in the MAGIX manual.

MAGIX framework is written in Python, while various algorithm packages are
written in Fortran for performance reasons. The following packages are required:
python~2.6, NumPy~1.3, pyfits~1.0, gfortran~4.3, matplotlib~0.99, and
scipy~0.9. The MAGIX package is a command-line-based program; therefore it can
be called by other programs, e.g.\ CASA\footnote{\url{http://casa.nrao.edu/}}.

    \section{Registration}\label{sec:registration}

In contrast to other astrophysical optimization packages MAGIX does not include
any intrinsic model program that calculates the model function depending on a
given parameter set. To communicate with the external model program, MAGIX has
to create the required input file(s) for every call of the external model
program, including the modified parameter values and a directive how to read in
the result of the model program. During every optimization step the values of
the parameters to be optimized will have been modified. Consequently, MAGIX has
to produce the actual input files that contain the new parameter values for the
model to run at each subsequent function call. Therefore MAGIX has to be given
directives how to create/write the input files that will be used in the function
call. Hence the user has to define the structure of the input file(s), including
loops with the information which parameter has to be written to which location.
After each optimization step, MAGIX compares the experimental data with the
values of the model function calculated by the external program with the latest
values of the parameters. The difference between data and model is quantified by
the value of $\chi^2$, i.e., low values of $\chi^2$ correspond to small
differences. Therefore, the path, the file name(s), and the format of the output
file(s) of the model program have to be defined as well. The descriptions of
both the input and the output files of the model are given in the XML file that
we call registration file.

Additionally, the registration file indicates whether the external model program
can be used in a parallelized MAGIX run or not, i.e., if it is possible to
execute two or more instances of the same external model program on the same
machine at the same time. This depends on how and where output or intermediate
files of the model are written, since different instances of models running in
parallel must not overwrite each other's files. This is mainly a bookkeeping
problem.

Ideally, a model has to be registered only once, i.e., it is not necessary to
register an already registered model again as long as the structure of the input
and output file(s) is unchanged. Whenever one wants to optimize some
parameter(s) of the model with MAGIX, it should be sufficient to edit the
instance XML file.

MAGIX comes with a suite of pre-registered models\footnote{At the moment the
following software packages are registered: SimLine, Lime, Radmc-3D, myXCLASS,
RADEX}, and the authors are willing to assist with the registration of new
models.

    \begin{figure}[t]
      \centering
      \includegraphics[width=0.48\textwidth]{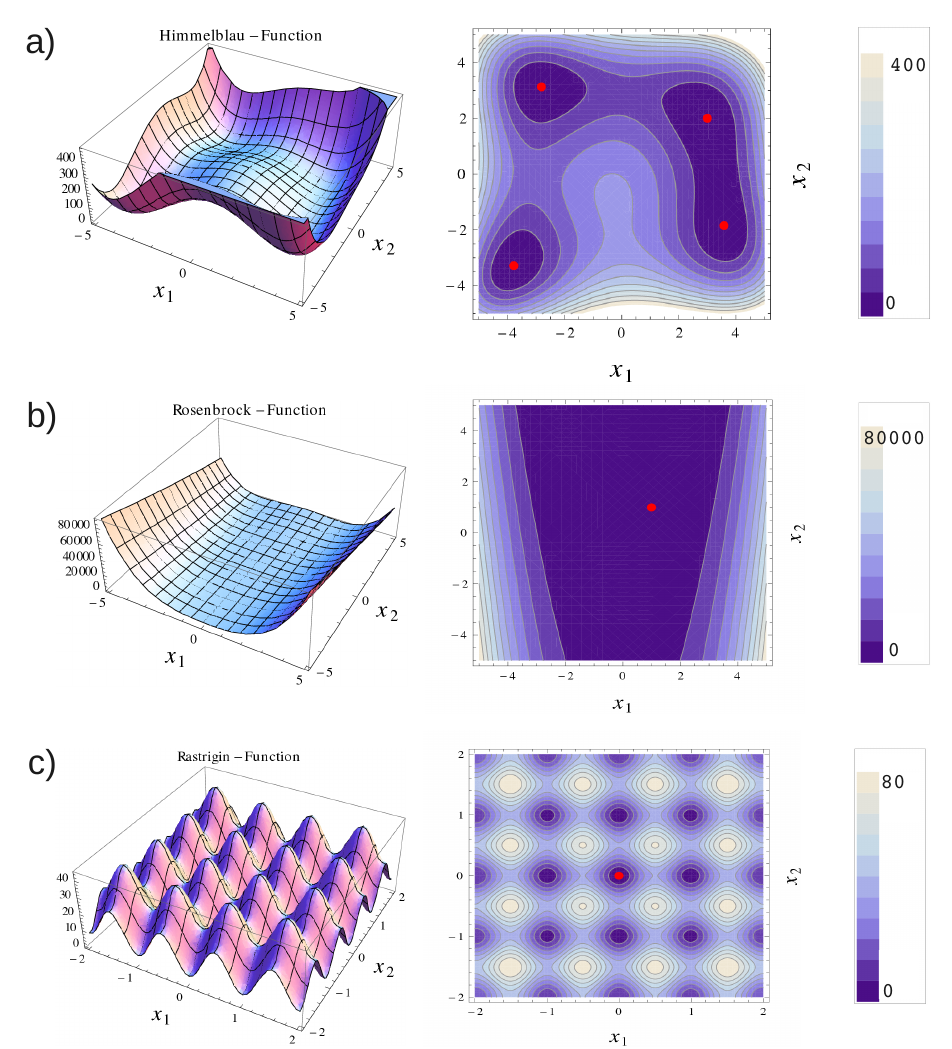}\\
      \caption{Analytical test functions: a)~The Himmelblau function $f(x_1,
x_2) = (x_1^2 + x_2 - 11)^2  + (x_1 + x_2^2 - 7)^2$ has four identical minima at
$f(3.0, 2.0) = f(-2.805118, 3.131312) = f(-3.779310, -3.283186) = f(3.584428,
-1.848126) = 0$. b)~The Rosenbrock function $f(x_1, x_2) = (1 - x_1)^2 + 100 \,
(x_2 - x_1^2)^2$ has one global minimum at $f(1, 1) = 0$. c)~The
Rastrigin function  $f(x_1, x_2) = 10 + (x_1^2 - 10 \cos(2 \pi \cdot x_1))
+ 10 + (x_2^2 - 10 \cos(2 \pi \cdot x_2))$ has one global minimum at $f(0, 0) =
0$. The contour plots of the different test functions are plotted in the second
column. The positions of the global minima for each test function are indicated
by red dots.}
      \label{fig:Testfunctions}
    \end{figure}

    \section{Optimization algorithms}\label{sec:optimization}

MAGIX provides optimization through one of the following algorithms or via a
combination of several of them (algorithm chain, see \S\ref{subsec:algchain}):
the Levenberg-Marquardt (conjugate gradient) method (\S\ref{subsec:LM}), which
is fast, but can get stuck in local minima, simulated annealing
(\S\ref{subsec:SA}) and particle swarm optimization methods
(\S\ref{subsec:PSO}), which are slower, but more robust against local minima.
Other, more modern methods, such as bees (\S\ref{subsec:Bees}), genetic
(\S\ref{subsec:GA}), nested sampling (\S\ref{subsec:NS}), or interval nested
sampling algorithms (\S\ref{subsec:INS}) are included as well for exploring the
solution landscape, checking for the existence of multiple solutions, and giving
confidence ranges. Additionally, MAGIX provides an interface to make several
algorithms included in the \texttt{scipy} \footnote{\url{http://www.scipy.org/}}
package available.

In the following, we give a short description of the optimization algorithms
that are implemented in MAGIX using the analytic Himmelblau-, Rosenbrock-, and
Rastrigin functions (Fig.~\ref{fig:Testfunctions}) as test functions for
demonstration, with multiple minima (Himmelblau, Rastrigin) and a very shallow
minimum (Rosenbrock). The optimization problem is solved directly through these
algorithms via stochastic searching without derivatives or gradient
information (except for the Levenberg-Marquardt algorithm).

    \subsection{Levenberg--Marquardt algorithm}\label{subsec:LM}

    \begin{figure}[t]
      \centering
      \includegraphics[width=0.49\textwidth]{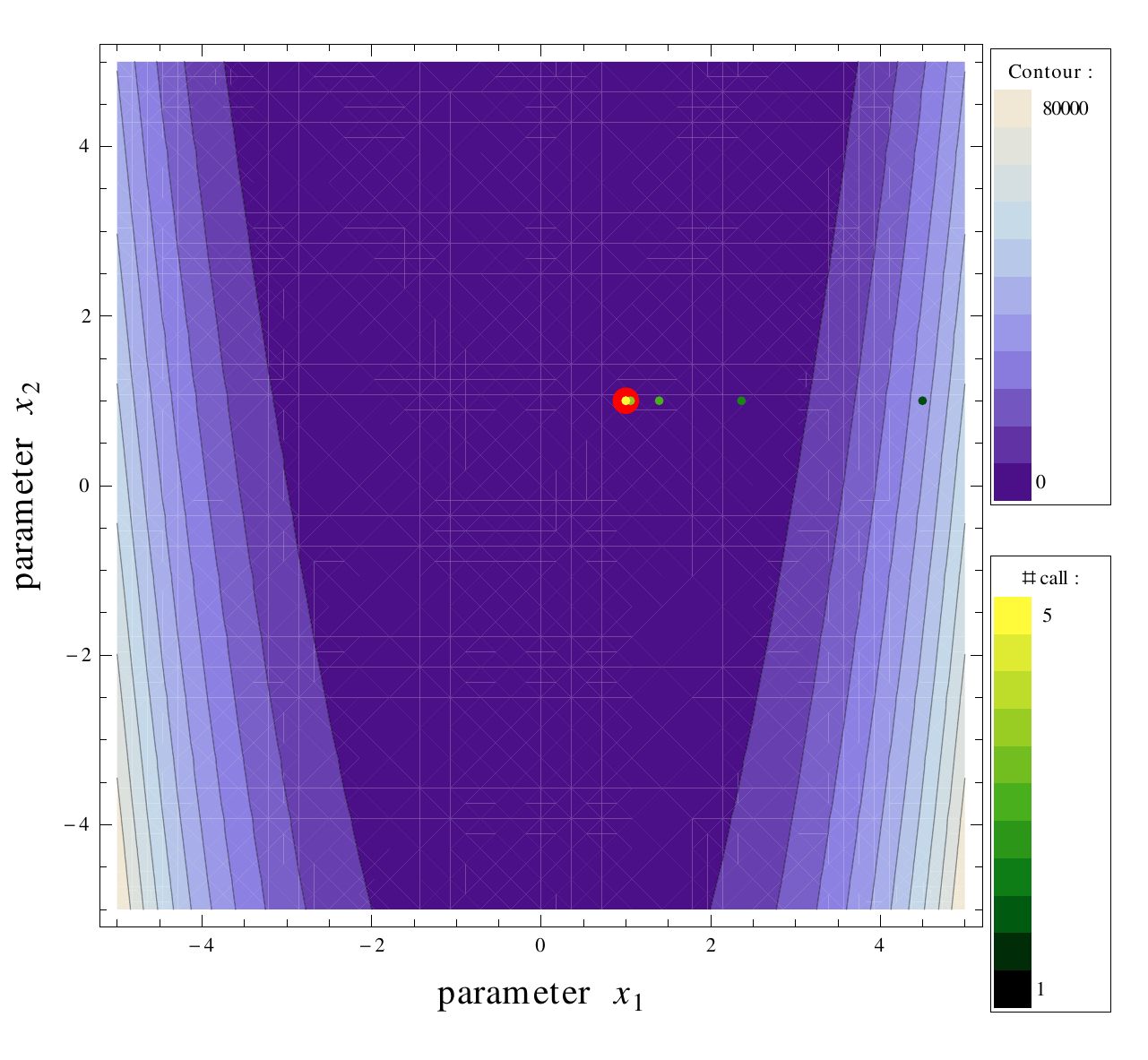}\\
      \caption{Results for the Rosenbrock function using the LM algorithm with
start values $x_1 = 4.5$, $x_2 = 1.0$: The distribution of the parameter values
after five function calls ($\chi^2=1.38 \cdot 10^{-9}$) is indicated by
green-yellow points. The sequence of iterations is color-coded (color bar on the
lower right). Early points are denoted in dark green, points toward the end of
the iteration are light yellow. The red dot denotes the global minimum of the
Rosenbrock function.}
      \label{fig:LM}
    \end{figure}

The Levenberg--Marquardt algorithm (LM), \citet{LMMarquardt, LMNocedal,
NRPreTeu07}\footnote{While we often refer to \textit{Numerical Recipes} (NR) to
explain the algorithms, no actual NR algorithms are included in the
package.} is a hybrid between the Gauss--Newton algorithm and the method of
gradient descent. The Gauss--Newton algorithm is a method for solving non-linear
least-squares problems. It is a modification of Newton's method to find the
minimum of a function, but is constrained so that it can only minimize a sum of
square function values. It requires knowledge of the gradients in $\chi^2$
space, which can be obtained from differential steps for sufficiently smooth
functions. The LM can find a minimum (possibly local) even if it starts very far
from it, but the efficiency depends on the landscape of the parameters. On the
other hand, for functions and starting values of parameters that are very close
to the final minimum, the LM tends to be slower than Gauss--Newton. The LM is an
algorithm that strongly depends on the starting values of the parameters that
are to be optimized, and the user should choose the starting values very
carefully. Otherwise the algorithm can easily become stuck in a side minimum of
the global solution.

MAGIX contains a modified version of the MINPACK package implementation
\citet{LMMinpack}. The gradient of the $\chi^2$ function is calculated in a
parallel environment using \texttt{OpenMP}. Furthermore, the user can define the
variation $var$ used for the gradient determination. Because MAGIX cannot
determine the gradient analytically, MAGIX has to use a numerical approximation:
$(\partial / \partial x_i) f(\vec{x}) = (f(x_i + h) - f(x_i))/h$, where the
variation $h$ is defined by $h = x_i \cdot var$. Varying the size of the
variation may be necessary if the $\chi^2$ function is not a smooth function and
the calculation of the gradient produces awkward results. As shown in
Fig.~\ref{fig:LM}, the fast convergence of the LM is obvious, where the minimum
is found after five (!) iterations.

    \subsection{Simulated annealing}\label{subsec:SA}

    \begin{figure}
      \centering
      \includegraphics[width=0.49\textwidth]{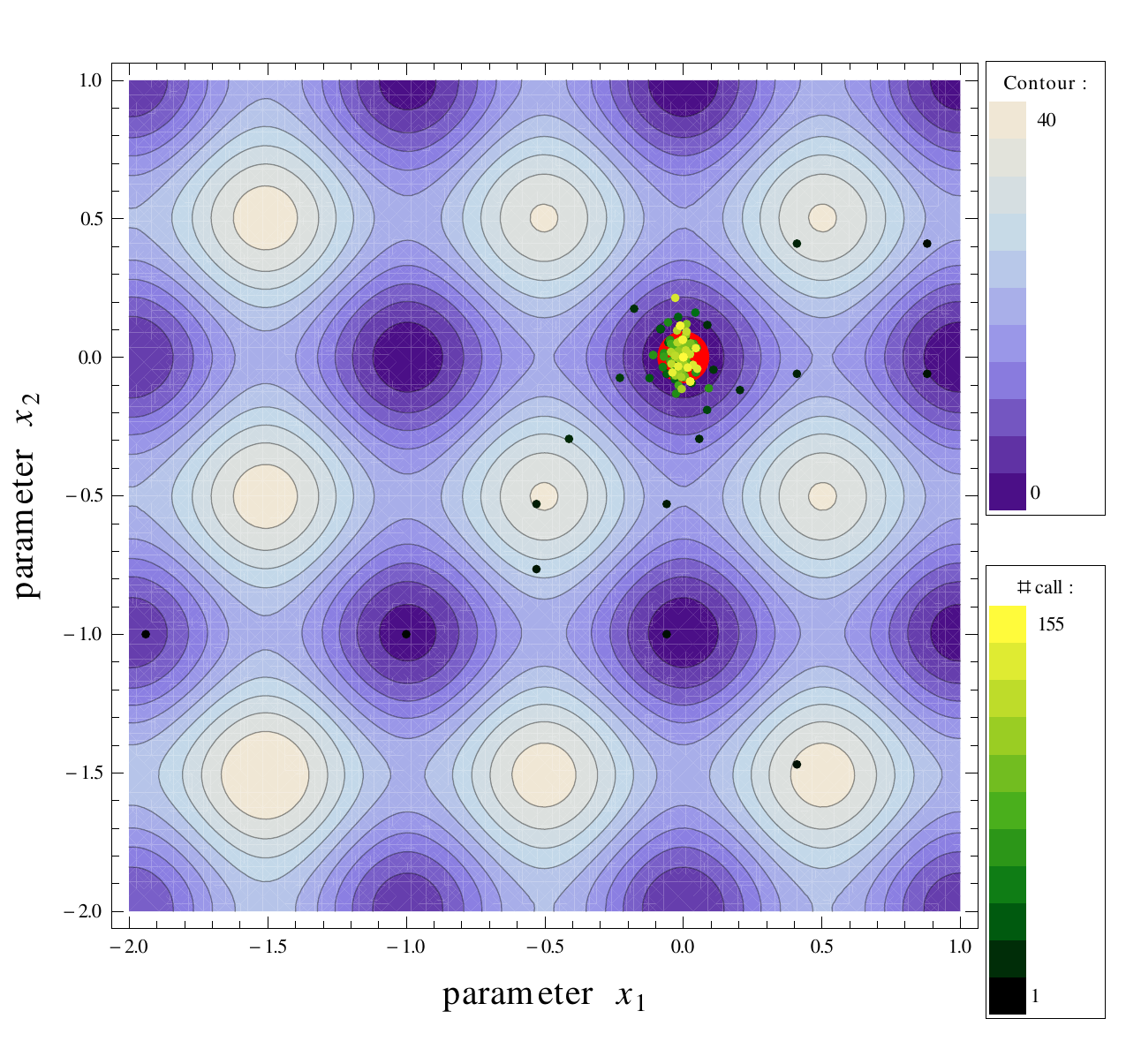}\\
      \caption{Results for the Rastrigin function using the SA algorithm with
start values $x_1 = -1.0$, $x_2 = -1.0$: The distribution of the parameter
values after 155 function calls ($\chi^2=4.37 \cdot 10^{-7}$) is indicated by
green-yellow points. The sequence of iterations is color-coded (color bar on the
lower right). Early points are denoted in dark green, points toward the end of
the iteration are light yellow. The red dot denotes the global minimum of the
Rastrigin function.}
      \label{fig:SA}
    \end{figure}

Simulated annealing (SA), \citet{NRPreTeu07} is a generic probabilistic
computational method that is used for the problem of global optimization,
i.e., to find a good approximation to the global optimum of a given function in
a large parameter space. For certain problems, SA is more effective than
exhaustive enumeration -- provided that the goal is merely to find an acceptably
good solution in a fixed amount of time, rather than the best possible solution.

In comparison to the LM, SA is more robust. Its result does not depend so much
on the neighborhood of the starting point. The LM searches for the highest
(negative) gradient and stops when it detects a local minimum. In contrast, SA
can check if the gradient around a local minimum is flat (low perturbation),
in which case it will continue to find a better minimum, i.e., a lower value
than the one found before. In Fig.~\ref{fig:SA}, the SA algorithm is able to
find the global minimum of the Rastrigin function at $x_1=0$ and $x_2=0$
although it starts in a local minimum at $x_1=-1$ and $x_2=-1$.

The name and inspiration of this algorithm come from the process of annealing in
metallurgy. This technique involves heating and controlled cooling of a
material, with the aim to gradually increase the size of its crystals and reduce
their defects. The heat provides energy and causes the atoms to move from their
initial position (which was a local minimum of internal energy) and wander
randomly through states of higher energy. Then, a slow cooling gives them more
chances of finding configurations of lower internal energy than the initial one.

By analogy to the physical process, each step of SA replaces the current
solution by a random nearby solution. This nearby solution is chosen with a
probability that depends both on the difference between the corresponding
function values and on a global temperature, $T$. The temperature $T$ is
gradually decreased (multiplied by the temperature reduction coefficient, $k <
1$) during the process.

The dependency of the temperature difference between two subsequent steps is
such that the solution changes almost randomly when $T$ is high, but it is
modified increasingly toward lower values as $T$ becomes zero. Allowing for
increasing values prevents the method from becoming stuck at local minima --
which can happen with gradient methods such as the LM. Subsequent points of the
SA algorithm follow a perpendicular direction.

MAGIX uses a partially parallelized implementation of the \texttt{scipy}
algorithm using \texttt{OpenMP}.

    \subsection{Particle swarm optimization}\label{subsec:PSO}

    \begin{figure}[t]
      \centering
      \includegraphics[width=0.49\textwidth]{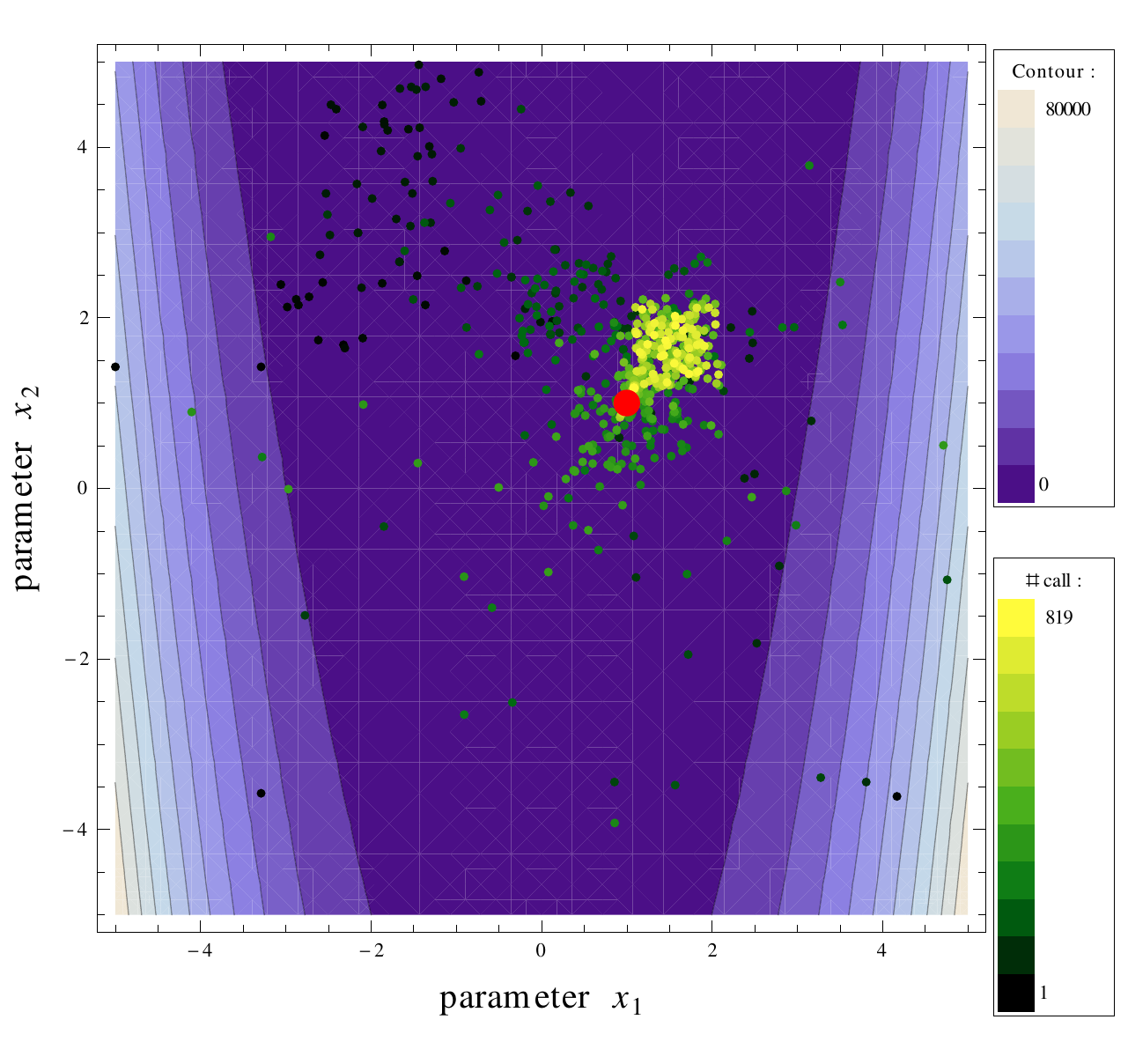}\\
      \caption{Results for the Rosenbrock function using the PSO algorithm: The
distribution of the parameter values after 819 function calls ($\chi^2=2.31
\cdot 10^{-5}$) is indicated by green-yellow points. The sequence of iterations
is color-coded (color bar on the lower right). Early points are denoted in
dark green, points toward the end of the iteration are light yellow. The red dot
denotes the global minimum of the Rosenbrock function. (The clustering of the
function calls right above the global minimum is caused by the very flat
gradient of the Rosenbrock function in this area.)}
      \label{fig:PSO}
    \end{figure}

The particle swarm optimization (PSO) algorithm implemented in MAGIX is a hybrid
\citet{PSOFan} between a particle swarm optimization algorithm
\citet{PSOKennedy} and a Nelder--Mead simplex search method \citet{PSONelder}.
The PSO optimizes a problem by iteratively trying to improve a candidate
solution according to some measure of quality; the particles are sent toward a
better solution flying so much faster according to the technique's performance
in previous step. The Nelder--Mead technique or downhill simplex search method
is a traditional technique for the direct search of function minima; it is easy
to use, does not need calculation of derivatives and therefore can converge even
to non-stationary solutions.

Both techniques have been modified by Fan and Zahara \citet{PSOFan} and their
combination is the hybrid algorithm that is available in MAGIX. The PSO
algorithm performs a kind of heuristics: It starts from an initial random
population of particles and searches in the neighborhood for a global optimum.
The particles with the best-fit function values are updated with the simplex
method, while the particles with the poorest function values are updated with
the PSO.

The whole procedure prevents the algorithm from being trapped locally and at
the same time allows it to find the global minimum. It repeats itself until a
termination criterion or the maximum number of iterations is reached. The
iteration shown in Fig.~\ref{fig:PSO} stops after 819 function calls because
the value of $\chi^2$ dropped below the limit of $\chi^2 = 9 \cdot 10^{-5}$. The
corresponding parameter values are $x_1=0.9954$, and $x_2=0.9907$ instead of
$x_1=1$ and $x_2=1$. Reducing the limit of the $\chi^2$ value would lead to
a better description of the global minimum but it would require more function
calls. The algorithm implemented in MAGIX is parallelized using
\texttt{OpenMP}.

    \subsection{Bees algorithm}\label{subsec:Bees}

    \begin{figure}[t]
      \centering
      \includegraphics[width=0.49\textwidth]{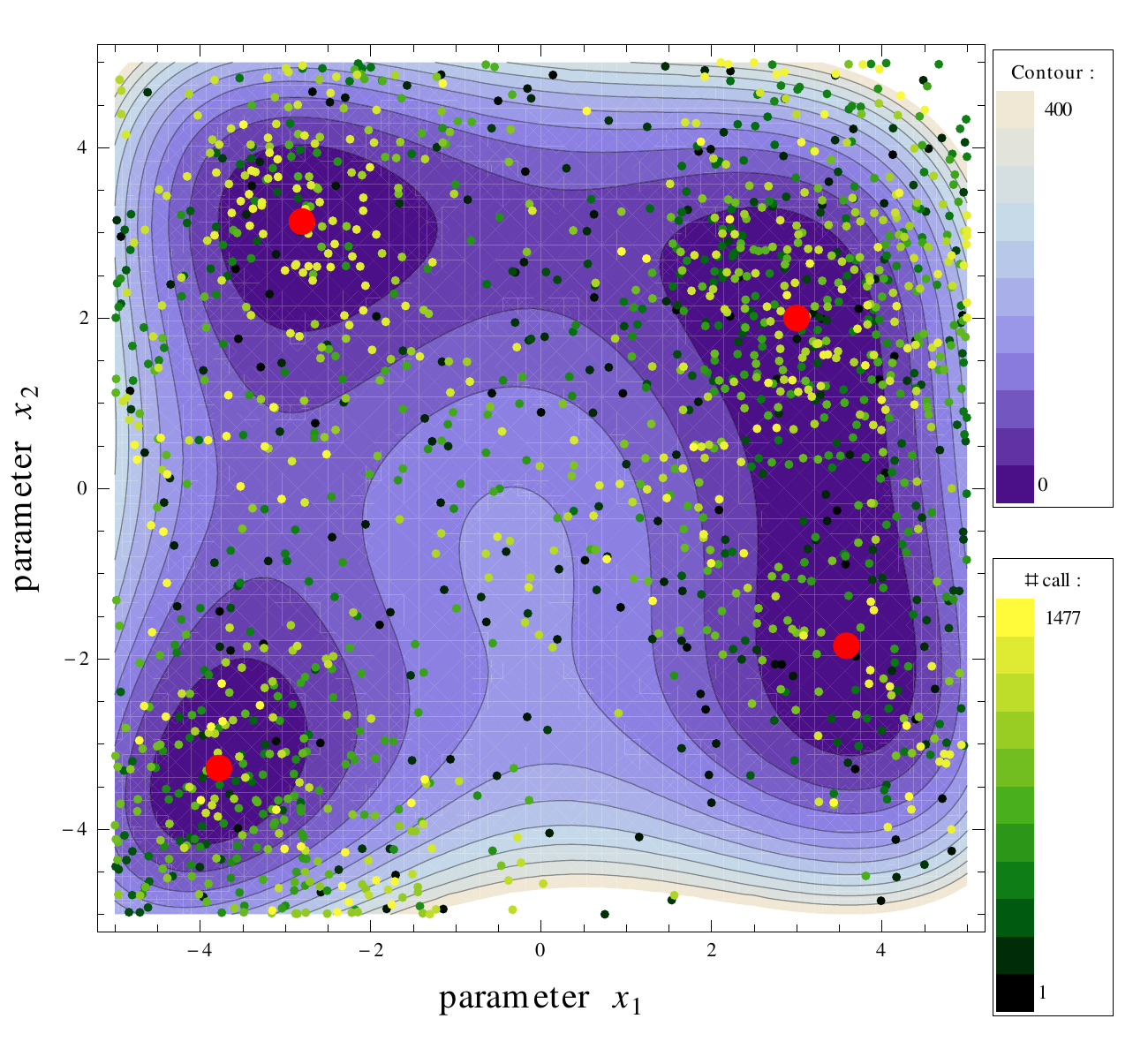}\\
      \caption{Results for the Himmelblau function using the bees algorithm: The
distribution of the parameter values after 1477 function calls is indicated by
green-yellow points. The sequence of iterations is color-coded (color bar on the
lower right). Early points are denoted in dark green, points toward the end of
the iteration are light yellow. The red dots denote the four global minima of
the Himmelblau function.}
      \label{fig:Bees}
    \end{figure}

The bees algorithm \citet{BeesPham} is a swarm algorithm that performs a kind
of neighborhood search combined with a random search (see, Fig.~\ref{fig:Bees}).
This name was given to the algorithm because it tries to mimic the collection of
nutriments by honey bees, in that there is always a part of the population that
performs the role of scouts, traveling far away in random directions to detect
new nutrition sources.

The algorithm starts with an initial set of parameter vectors (a collection of
particles or scout bees, i.e., the hive of bees), randomly selected and such
that it spreads throughout the entire parameter space. After the fitness of each
bee is evaluated in terms of the quality ranking it just visited, the bees
with the highest fitness visit the neighborhoods of the sites they currently
are at. Of the remaining bees some are sent away to random sites and some
are sent to search in the neighborhood of the very best sites as well (so that
there are more bees searching for food in places where it is more probable to
find nutrition sources). At the end of the step, the fitness of each visited
site is evaluated, and the bees who just visited them move away or search in the
closer vicinity of the best sites. The result is that the bees algorithm finds
areas of local minima, see Fig.~\ref{fig:Bees}.

    \subsection{Genetic algorithm}\label{subsec:GA}

    \begin{figure}[t]
      \centering

\includegraphics[width=0.49\textwidth]{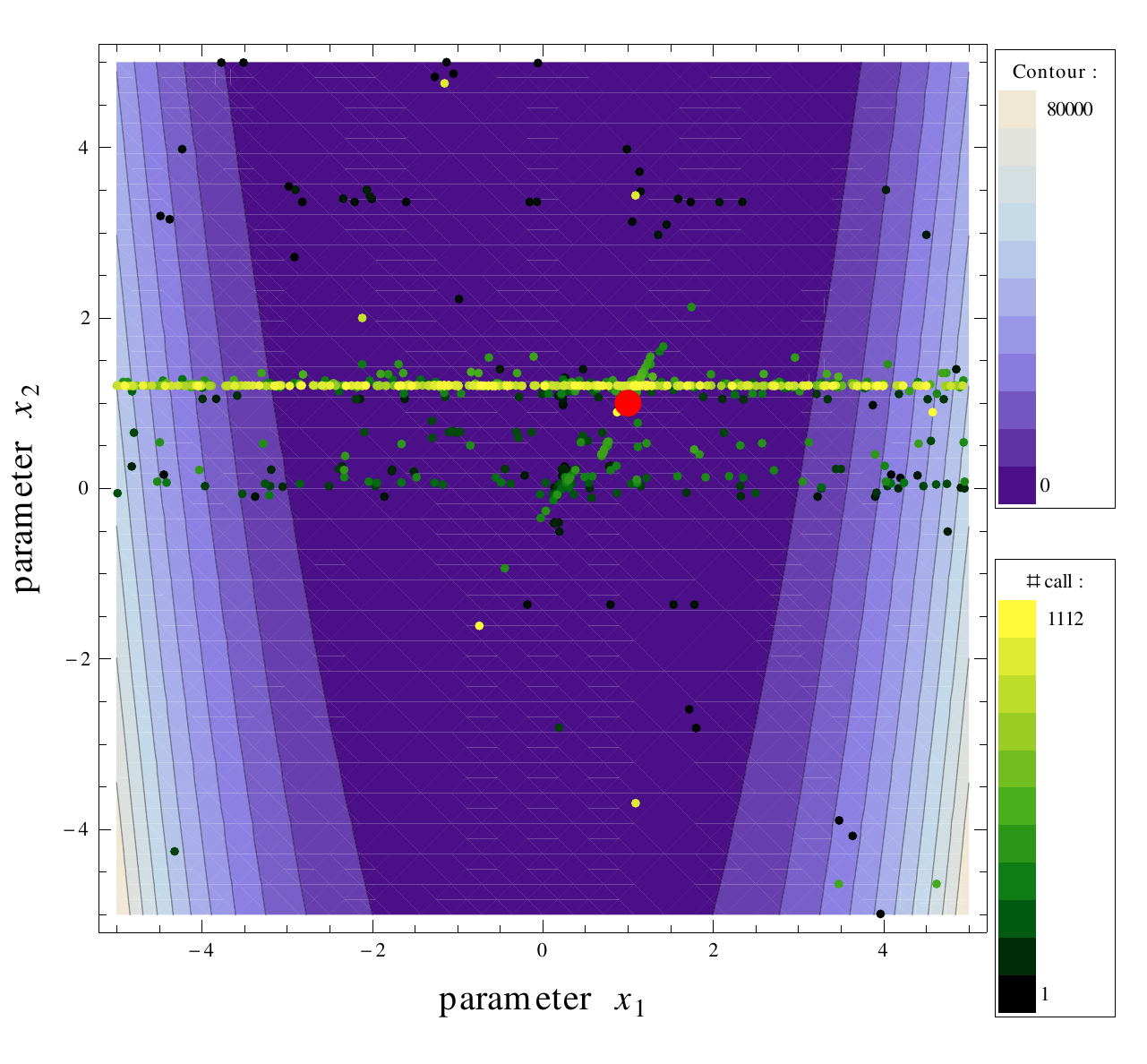}\\
      \caption{Results for the Rosenbrock function using the GA: The
distribution of the parameter values after 1112 function calls ($\chi^2=9.33
\cdot 10^{-3}$) is indicated by green-yellow points, where the dark green points
indicate function calls that are made at the beginning of the fit process and
light yellow points represent function calls at the end of the fit process. The
red dot denotes the global minimum of the Rosenbrock function.}
      \label{fig:GA}
    \end{figure}

The genetic algorithm (GA) is a probabilistic search algorithm that mimics the
process of natural evolution. It iteratively transforms a set of parameter
vectors (population), each with an associated fitness value, into a new
population of objects.

The procedure takes place using the Darwinian principle of natural selection,
with operations that are patterned after naturally occurring genetic operations
such as recombination and mutation. Recombination is the joined process of
reproduction and crossover, i.e., mixing the genetic matter (parameter values)
of the parents (parameter vector) \citet{GAWhitley}. Mutation is the complete
disappearance of specific genetic material and its transformation to a
completely different material; this happens especially in combinations of
genetic material (parameter sets) that does not fit.

The evolution starts from a randomly selected population (of parameter sets) and
proceeds in evolutionary generations (stages/iteration steps). When a generation
step is completed, the fitness of all members of the population is evaluated.
Then a number of parameter sets is stochastically selected for modification; the
fittest members are more likely to be kept unmodified. The rest of the
population members are modified; the modifications they go through are more
intense because they are fit. The modified and unmodified parameter sets make up
the new population whose fitness will be evaluated at the end of the step
\citet{GAHerrera1, GAHerrera2} (see, Fig.~\ref{fig:GA}).

    \subsection{Nested sampling}\label{subsec:NS}

    \begin{figure}[ht]
      \centering
      \includegraphics[width=0.49\textwidth]{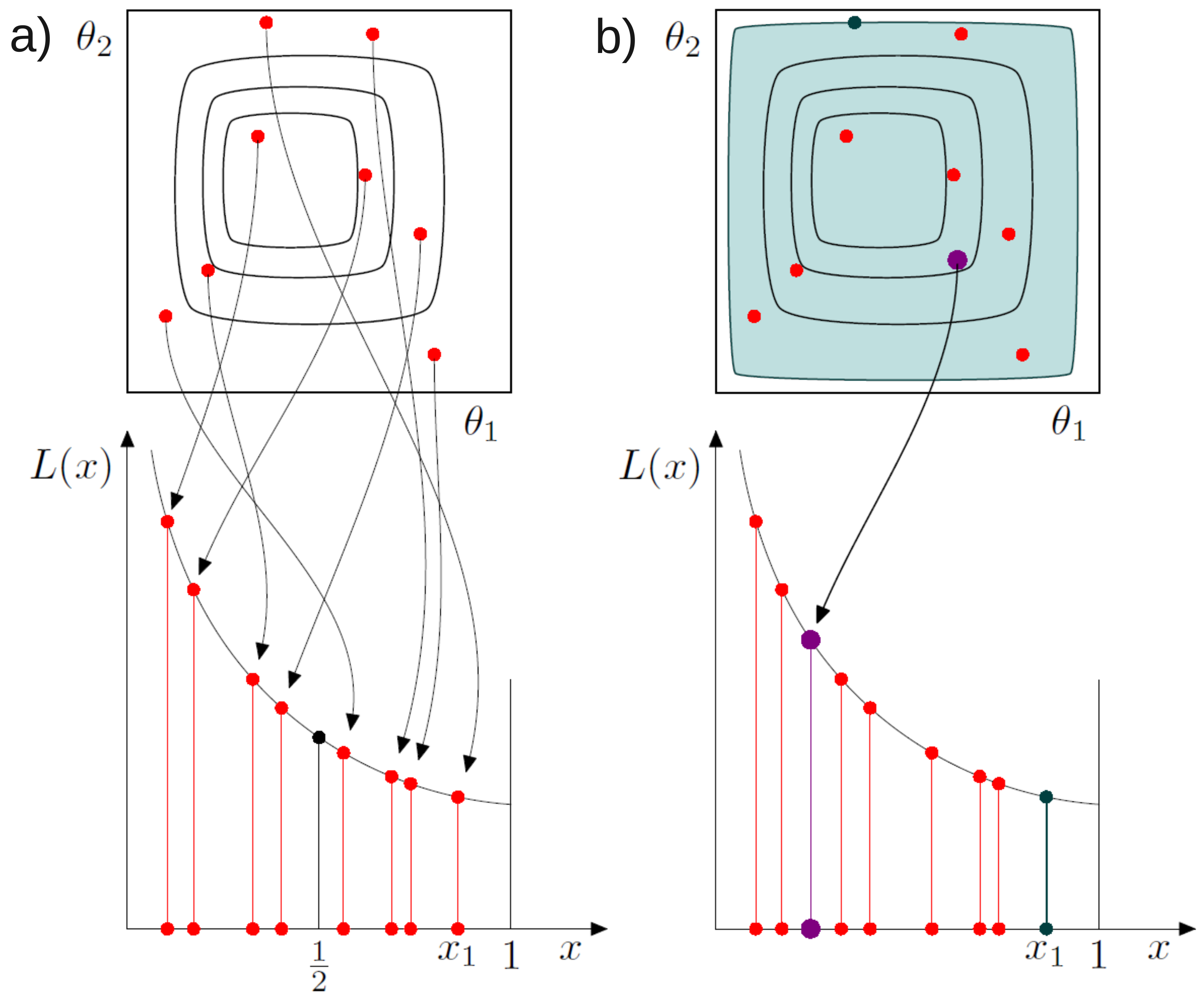}\\
      \caption{Principle of the NS algorithm (taken from
\citet{NSSkillingMacKay}): The upper part of both panels describes a contour
plot of a likelihood function $\mathcal{L}(\theta)$. Left panel a): Each point
$\theta$ within parameter space $\wp$ defined by the parameter ranges of
$\theta_1$ and $\theta_2$ is associated with the volume that would be enclosed
by the contour $L=\mathcal{L}(\theta)$. ($L(x)$ is the contour value such that
the volume enclosed is $x$). If the points $\theta$ are uniformly distributed
under the prior probability distribution (prior), all these volumes ($x$-values)
are uniformly distributed between 0 and 1. Right panel b): Using a Markov chain
method, the NS algorithm takes a point (purple dot) from $\wp$ satisfying $L \ge
L(x_1)$. Inserting the new point into this distribution, we can find the highest
$x$-value $x_2$ used for the next iteration.}
      \label{fig:NSBasic}
    \end{figure}

The~nested~sampling~(NS) \citet{NASkilling, NAFeroz} algorithm included in MAGIX
is a combination of a Monte Carlo method and Bayesian
statistics~\citet{NSSivia}. The Monte Carlo methods are a family of
computational algorithms that perform repeated random sampling and compute the
results for every sample. The Bayesian statistics is a statistical inference
technique, i.e., it attempts to draw conclusions from data subject to random
variation. In particular, the Bayesian inference calculates the probability that
a hypothesis is true, i.e., how probable it is that the newly calculated value
of a parameter is closer to the real one (the value that best fits experimental
data); if it is more probable than the previously calculated, then the parameter
value is updated.

The principle of the NS algorithm is illustrated in Fig.~\ref{fig:NSBasic}: The
Bayesian approach is used to obtain a set of physical parameters $\Theta =
(\theta_1; \theta_2, \ldots, \theta_n)$ that attempts to describe the
experimental data. The Bayesian analysis is assumed to incorporate the prior
knowledge with a given set of current observations to make statistical
inferences. The prior information $\pi (\Theta)$ can come from observational
data or from previous experiments. Bayes theorem states that the posterior
probability distribution of the model parameters is given by

\begin{equation}
  {\rm Pr}(\Theta) = \frac{\mathcal{L}(\Theta) \, \pi(\Theta)}{Z},
\end{equation}

where Pr$(\Theta)$ is the posterior probability distribution of the model
parameters, $\mathcal{L}(\Theta)$ is the likelihood of the data for the given
model and its parameters, $\pi(\Theta)$ is a prior information, and $Z$ is
Bayesian evidence. The Bayesian evidence is the average likelihood of the model
in parameter space. It is given by the following integral over the
$n$-dimensional space:

\begin{equation}\label{NS:Z}
  Z = \int \mathcal{L}(\Theta) \pi(\Theta) \, d\Theta.
\end{equation}

The NS algorithm transforms the integral (\ref{NS:Z}) to the single dimension by
re-parametrization to a new linear variable\footnote{Although we do not know the
values of these volumes $X$, we know the order of them because $L(x_i)=
\mathcal{L}(\theta)$} -- a {\it prior volume} $X$. The volume of parameter space
can be divided into elements $dX = \pi(\Theta) d\Theta$. The prior volume $X$
can be accumulated from its elements $dX$ in any order, so we construct it as a
function of decreasing likelihood:

\begin{equation}
  X(\lambda) = \int_{\mathcal{L}(\Theta) > \lambda} \pi(\Theta) \, d\Theta.
\end{equation}

That means that the cumulative prior volume covers all likelihood values
exceeding $\lambda$. As $\lambda$ increases, the enclosed volume $X$ decreases
from $X(0) = 1$ to $X(1) = 0$. If the prior information $\pi(\Theta)$
is uniformly distributed in the parameter space, equation (\ref{NS:Z})
for the evidence transforms into

\begin{equation}
  Z = \int_0^1 \mathcal{L}(X) \, d X.
\end{equation}

One can calculate the partial likelihood as $L_i = \mathcal{L}(X_i)$, where the
$X_i$ is a sequence of decreasing values, such that

\begin{equation}
  0 < X_m < \ldots < X_2 < X_1 < 1.
\end{equation}

MAGIX uses the trapezoid rule to approximate the evidence

\begin{equation}\label{NS:Zi}
  Z = \sum_{i=1}^m Z_i, \quad {\rm where} \quad Z_i = L_i \, \frac{X_{i-1} -
X_{i+1}}{2}.
\end{equation}

The NS algorithm is targeted at calculating Bayesian evidence, but it assists in
obtaining a posterior sample of points from which one can estimate uncertainties
of parameter values. The prior volume $X_i$, which corresponds to the likelihood
contour $L_i$, is usually evaluated with a random number generator.

The actual algorithm is based on the Markov chain Monte Carlo (MCMC) methods
\citet{NRPreTeu07, NSMCMC, NSDiaconis}. Those are a family of algorithms that
sample from probability distributions, with the aim to construct a (Markov)
chain with the desired distribution (a distribution with the desired properties)
as the equilibrium distribution. The quality (how well the parameter sets fit
the data) of the sample is a monotonically increasing function of the number of
steps (see, Fig.~\ref{fig:NSBasic}).

    \begin{figure}[t]
      \centering
      \includegraphics[width=0.49\textwidth]{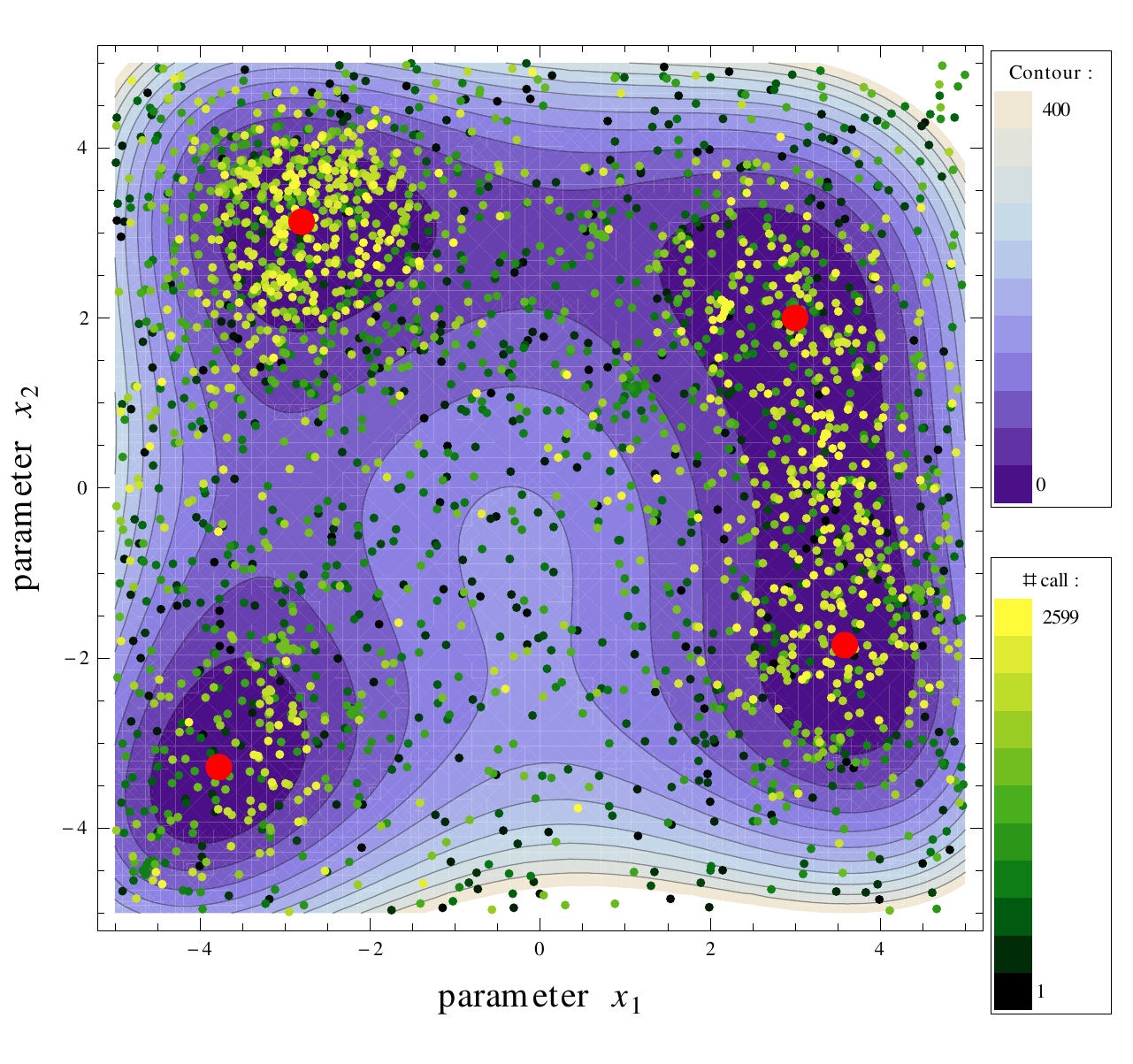}\\
      \caption{Results for the Himmelblau function using the NS
algorithm: The distribution of the parameter values after 2599 function calls is
indicated by green-yellow points, where the dark green points indicate function
calls which are made at the beginning of the fit process and light yellow points
represent function calls at the end of the fit process. The red dots denote the
four minima of the Himmelblau function.}
      \label{fig:NS}
    \end{figure}

    \begin{figure*}
      \centering
      \includegraphics[width=0.93\textwidth]{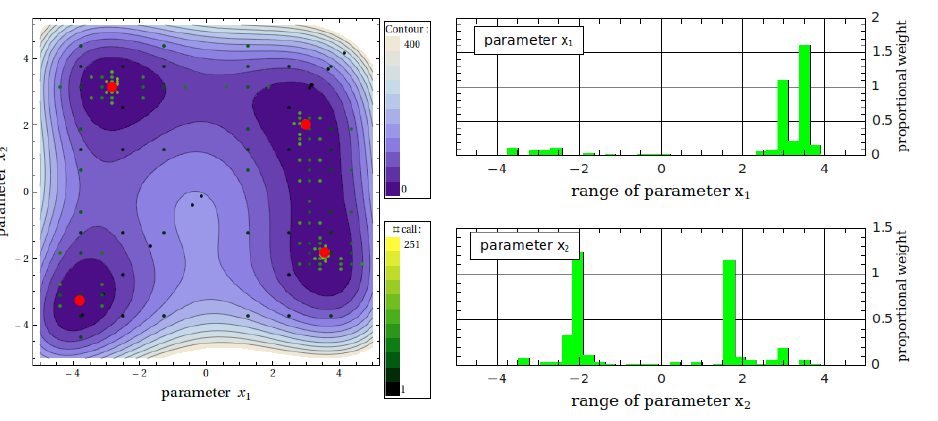}\\
      \caption{Results for the Himmelblau function using the INS algorithm: Left
panel: The distribution of the parameter values after 251 function calls is
indicated by green-yellow points, where the dark green points indicate function
calls that are made at the beginning of the fit process and light yellow points
represent function calls at the end of the fit process. The red dots denote the
four minima of the Himmelblau function. Right panel: Posterior weights of points
after the NS process.}
      \label{fig:INS}
    \end{figure*}

An MCMC algorithm is used to reduce the dimensionality of the parameter space
through integration. The Bayesian approach to this technique allows one not only
to find multiple solutions, but also to define proportional weights of parameter
values and to evaluate the Bayesian evidence\footnote{Bayesian interpretation of
probability: As the number of steps increases, we collect evidence with regard
to the consistency or inconsistency of that evidence with a given hypothesis.
More specifically, as the evidence accumulates, we tend to believe in the given
hypothesis more or less, depending on the degree to which the increasing
evidence agrees with that hypothesis.}. The NS algorithm included in MAGIX
requires fewer samples than standard MCMC methods \citet{NAFeroz} while also
providing posterior probabilities for each one of the best parameter vectors.
The NS algorithm has the following advantages: It is a non-derivative method and
investigates the landscape of optimization function. Additionally, it can find
multiple minima. On the other hand, the algorithm does not converge to a global
minimum and depends strongly on the random number generator. The results for the
Himmelblau function, shown in Fig.~\ref{fig:NS}, are similar to the results of
the bees algorithm. Both algorithms can be used to explore the landscape of the
$\chi^2$ distribution and for finding areas of global minima. The implementation
of the algorithm included in MAGIX is parallelized using \texttt{OpenMP}.

    \subsection{Interval nested sampling algorithm}\label{subsec:INS}

The Interval Nested Sampling (INS) algorithm (developed by I.~Bernst\footnote{A
paper describing the INS algorithm in detail is in preparation.}) included in
MAGIX is an implementation of the branch-and-bound algorithm \citet{INSIchida}
to find the next prior volume $X_i$. The algorithm is based on the NS algorithm
(\S\ref{subsec:NS}) and uses an interval method for the definition of a next
part of the prior volume. The main principle of the interval method is a
division of the parameter space into interval boxes and an estimation of the
optimization function value over the boxes. The estimate is called inclusion
function and can be calculated by various methods. The centered form of
inclusion with slopes is used in the current version of the INS algorithm
\citet{INSKrawczyk}. The interval method assists in finding the next prior
volume $X_i$ by determining the ratio of the volume of the working interval box
$Z_i$ to the whole volume $Z$ of the parameter space. Posterior weights of
points after the NS process are calculated using the Bayes theorem (see,
Fig.~\ref{fig:INS}):

\begin{equation}\label{INS:wi}
  w_i = \frac{Z_i}{Z}.
\end{equation}

Using the sequence of posterior samples of parameter vectors we are able to
determine the reliability of model parameters such as standard deviations or to
construct posterior distributions of parameter values. The mean value
$\mu(\theta_j)$ of each model parameter $\theta_j$, $j = (1, \ldots, n)$ and its
standard deviation $\sigma(\theta_j)$ are given as

\begin{equation}\label{INS:mean}
  \mu(\theta_j) = \sum_{i=1}^k w_i \, \theta_j
\end{equation}

and

\begin{equation}\label{INS:stdev}
  \sigma(\theta_j) = \sqrt{ \sum_{i=1}^k w_i \left( \theta_j - \mu(\theta_j)
\right)^2},
\end{equation}

where $k$ indicates the number of points in the sample. The INS algorithm is
capable of handling multi-modality of the optimization function, phase
transitions, and strong correlations between model parameters. As shown in
Fig.~\ref{fig:INS}, the Interval Nested Sampling algorithm requires fewer
function calls than other global optimizers. Additionally, it is used to
determine the confidence intervals of parameters, which is described in the next
section.

    \subsection{Error estimation}

\begin{figure}[!b]
      \centering
      \includegraphics[width=0.5\textwidth]{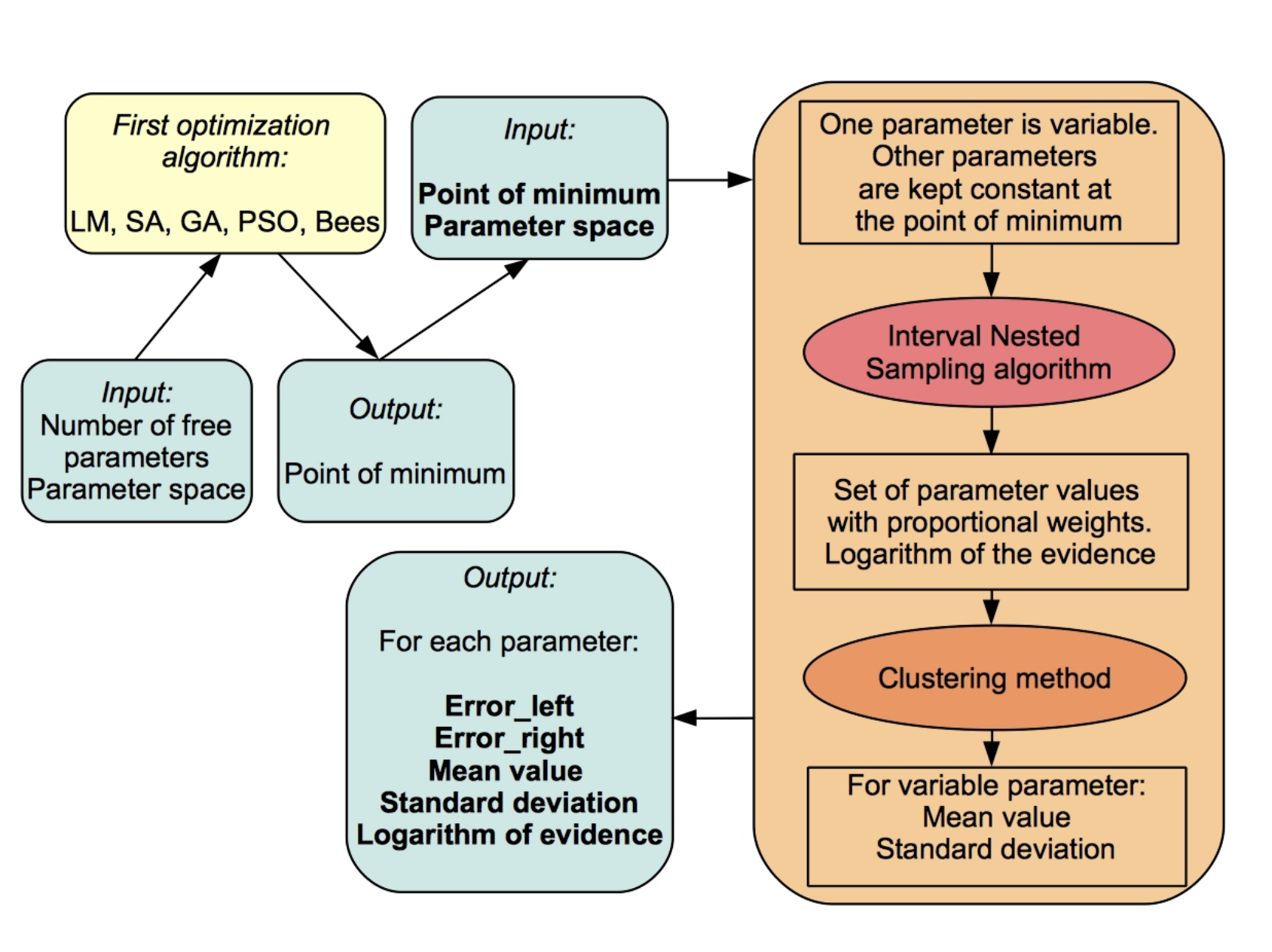}\\
      \caption{Schematic diagram of error estimation module of MAGIX.}
      \label{fig:ErrorEstimShema}
    \end{figure}

\begin{figure}[!b]
      \centering
      \includegraphics[width=0.5\textwidth]{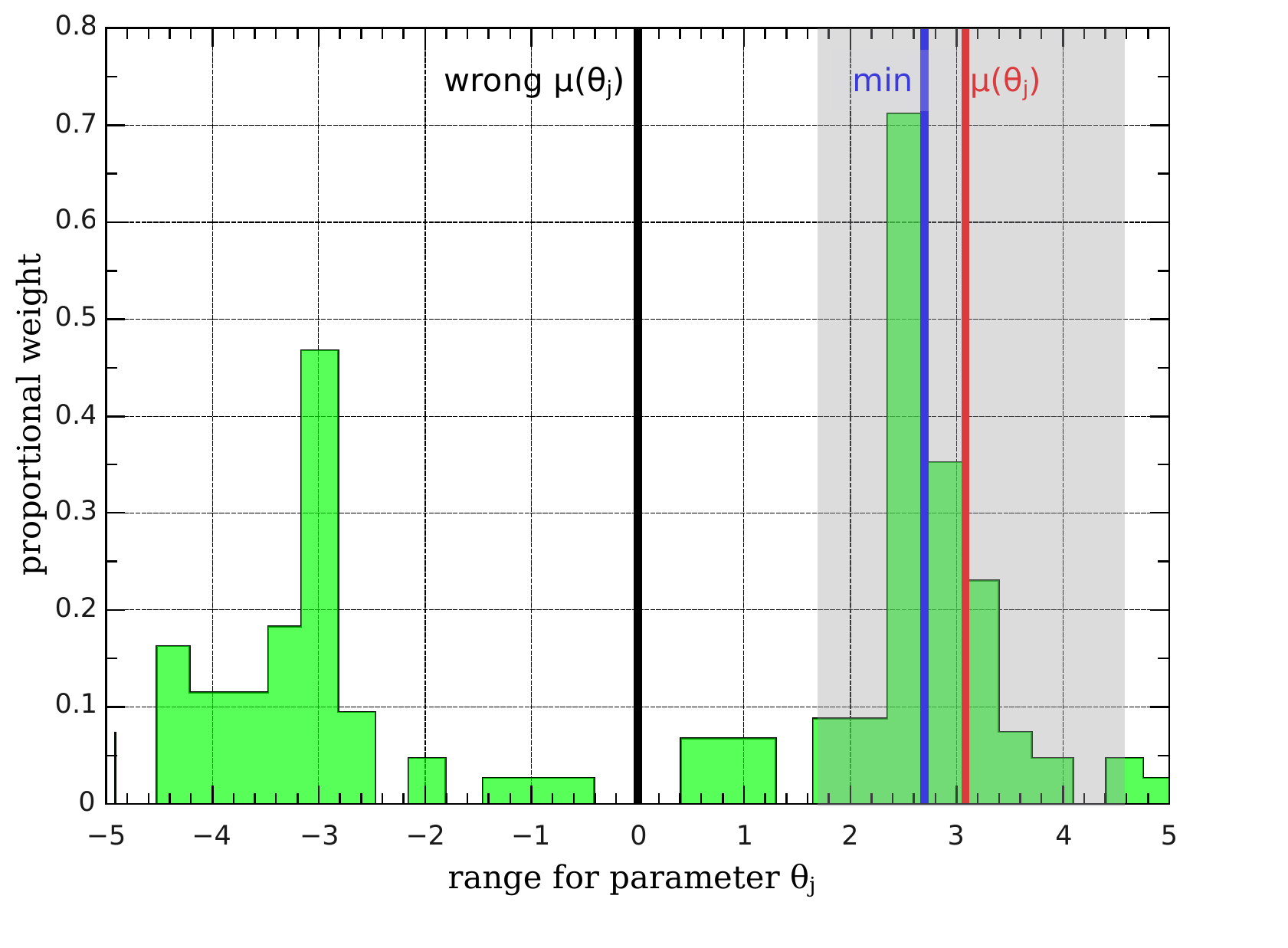}\\
      \caption{Distribution of parameter values with several minima where a
direct application of Eqn.~(\ref{INS:mean})~-~(\ref{INS:stdev}) is not possible.
(Here, $\mu(\theta_j)$ indicates the mean value $\mu(\theta_j)$,
$\sigma(\theta_j)$  is the standard deviation $\sigma(\theta_j)$, and ``min''
represents the value of the parameter $\theta_j$ of the best-fit result).}
      \label{fig:ErrorEstim_distr}
    \end{figure}

MAGIX provides an error estimation for each single parameter at the point of
minimum. A schematic diagram of the error estimation module is shown in
Fig.~\ref{fig:ErrorEstimShema}: After some optimization procedure MAGIX
determines a point of minimum. The input values for the error estimation are
given by the point of minimum and the parameter space. To determine the error of
a parameter $\theta_j$ at the minimum, MAGIX varies this parameter within the
given parameter range, while the other parameters are kept constant. The
INS algorithm is applied and returns a set of parameter values with proportional
weights and the logarithm of the Bayesian evidence for parameter $\theta_j$ for
a sequence of parameter values distributed over the whole parameter space. If
the distribution of parameter values has only one minimum,
Eqn.~(\ref{INS:mean})~-~(\ref{INS:stdev}) can be applied to calculate the mean
value and the standard deviation. Sometimes there are several minima in the
sequence, hence these formulas cannot be used directly because the resulting
estimation of the mean value and standard deviation produces meaningless results
(see Figs.~\ref{fig:INS}~and~~\ref{fig:ErrorEstim_distr}).

     \begin{figure}[t]
       \centering
       \includegraphics[width=0.5\textwidth]{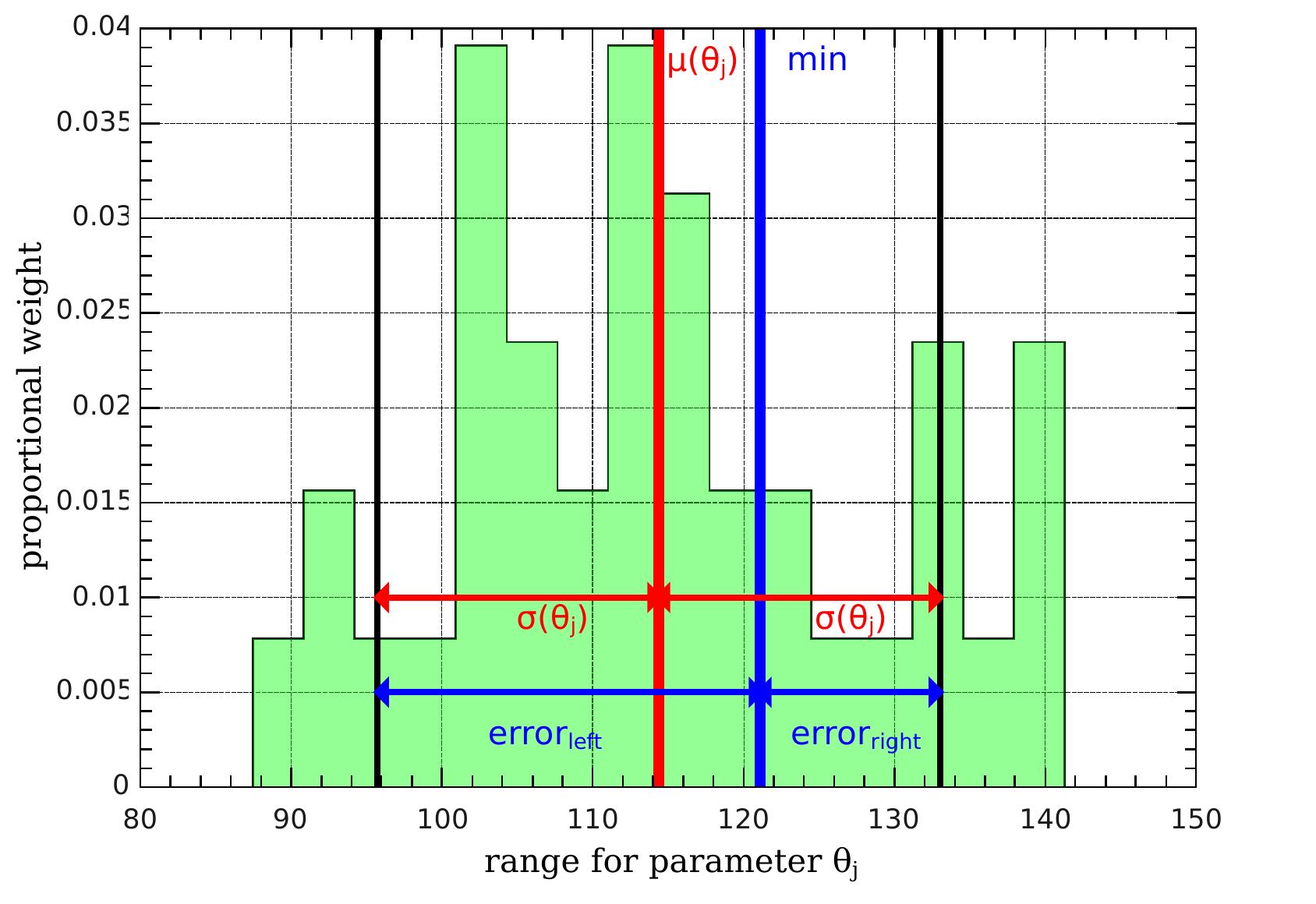}\\
       \caption{Schematic diagram of error estimation module of MAGIX. (Here,
$\mu(\theta_j)$ indicates the mean value $\mu(\theta_j)$, $\sigma(\theta_j)$
represents the standard deviation $\sigma(\theta_j)$, and ``min'' the value of
the parameter $\theta_j$ of the best-fit result).}
        \label{fig:ErrorEstim_hist}
     \end{figure}

We are interested in the uncertainty around the considered minimum. Therefore,
we need to estimate the mean value and the standard deviation in the minimum.
This is done using a clustering method:
\begin{itemize}
  \item Calculate the distances from the minimum to all points of the sample.
  \item Sort the points depending on their distances (ascending order).
  \item Select the points with a function value lower than the $\chi^2$
        boundaries for the 99 percent confidence region $\Delta \chi^2_{\alpha,
        n}$ ($\alpha$~=~0.99).
\end{itemize}
Finally, the new sample of $m$ points ($m < k$) is distributed around the
minimum point (Fig.~\ref{fig:ErrorEstim_distr}, gray box) and the mean value of
the parameter $\theta_j$ is calculated as follows:

\begin{equation}
  \mu \left(\theta_j \right) = \frac{\sum_{i=1}^m w_i \theta_j}{\sum_{i=1}^m
w_i}.
\end{equation}

In Bayesian statistics one usually applies a credible interval (or Bayesian
confidence interval) to the parameter value. For a single parameter and a sample
of points that can be summarized in a single sufficient statistic,
it can be shown \citet{INSJaynes} that the credible interval and the confidence
interval coincide if the unknown parameter is a location parameter (i.e.,
the forward probability function has the form $Pr$($\theta_i|\mu(\theta_i)$) =
$f$($\theta - \mu$) ) with a uniform prior distribution. Hence credible
intervals in our case are analogous to confidence intervals in frequentist
statistics. Using the mean value $\mu(\theta_j)$ and the standard deviation
$\sigma(\theta_j)$, the 3$\sigma$ confidence interval of the parameter
$\theta_j$ is given by

\begin{equation}
  Pr(\mu(\theta_j) - \sigma(\theta_j) < \theta_j < \mu(\theta_j) +
\sigma(\theta_j)) \approx 0.99.
\end{equation}

In Fig.~\ref{fig:ErrorEstim_hist} the example of a histogram of distribution of
parameter values after error estimation is shown. Hence for the parameter
value at the point of minimum (using the error left and the error right),

\begin{equation}
  Pr(\theta_i({\rm min}) - {\rm error}_{\rm left} < \theta_j < \theta_i({\rm
min}) + {\rm error}_{\rm right}) \approx 0.99.
\end{equation}

The Bayesian evidence usually plays an important role in model selection but in
parameter estimation the evidence factor is ignored because it is an integrated
value over the whole parameter space. The INS algorithm calculates the logarithm
of the evidence and can be used to estimate the quality of the fitting
procedure. A high absolute value of the evidence logarithm indicates a
big uncertainty of the parameters.

    \subsection{Which algorithm should be used?}

    \begin{figure*}[t]
      \centering
      \includegraphics[width=0.95\textwidth]{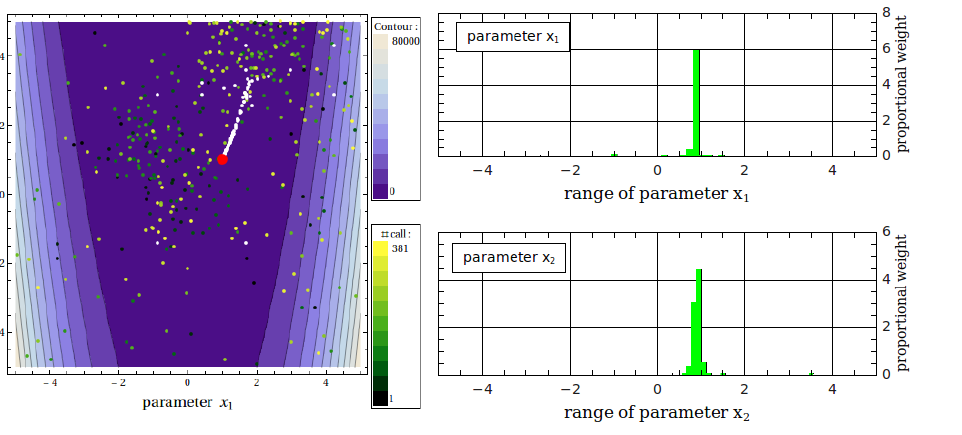}\\
      \caption{Results for the Rosenbrock function using an algorithm chain:
Left panel: The distribution of the parameter values for the bees algorithm
(green-yellow points, where the dark green points indicate function calls that
are made at the beginning of the fit process and light yellow points represent
function calls at the end of the fit process) and for the following simulated
annealing algorithm (white points). The red dot indicates the global minimum of
the Rosenbrock function. Right panel: Posterior weights of points after the
NS process used by the error estimation module.}
      \label{fig:AlgChain}
    \end{figure*}

In principle, it is impossible to answer the question which algorithm is best.
All algorithms included in MAGIX try to find the global minimum of the $\chi^2$
function, which depends on the observational data, on the external model, on the
free and fixed parameters, and on the ranges for each free parameter. Every
algorithm has a different strategy for finding the minimum of the $\chi^2$
function. Whether the strategy is successful depends on the $\chi^2$ function
and many other conditions. For example, the bees algorithm requires a huge
computational effort, but this algorithm explores the whole landscape of the
$\chi^2$ function within the given ranges. It gives a good overview of the
landscape of the problem and can in principle find even minima in narrow
valleys. On the other hand the computational effort is heavy, and the algorithm
is inefficient if computing the external model requires more than a few seconds.
Here, the computation time of an algorithm depends on the number of function
calls for each iteration, the computation time of the external model program,
the possibility of executing the external model program more than once on the
same machine at the same time and on the size of the parameter space. Other
swarm algorithms such as the particle swarm or NS algorithms might be
better if the $\chi^2$ function has a smoother shape. For example, the particle
swarm algorithm included in MAGIX makes use of a Nelder--Mead simplex search
method, which accelerates the whole computation if the $\chi^2$ function has no
narrow valleys, i.e, the $\chi^2$ function does not change drastically when one
or more parameters are varied.

\begin{table}[!b]
  \caption{Total cost (in function evaluations) for each test function for each
global optimization algorithm.}
  \centering
  \begin{tabular}{c|c|c|c}
    \hline
    algorithm & Rastrigin                & Rosenbrock                             & Himmelblau\\
    name      & function                 & function                               & function  \\
              & $\chi^2_{\rm limit} = 1$ & $\chi^2_{\rm limit} = 4 \cdot 10^{-3}$ & $\chi^2_{\rm limit} = 5 \cdot 10^{-4}$\\
    \hline
    Bees      & $1220$                   & $14491$                                & $101664$ \\
    PSO       & $1317$                   & $535$                                  & $770$ \\
    Genetic   & $241$                    & $533$                                  & $1626$ \\
    NS        & $4230$                   & $5080$                                 & $8720$ \\
    INS       & $20$                     & $1144$                                 & $168$ \\
    \hline
    \hline
  \end{tabular}
  \tablefoot{The algorithms stop if the value of $\chi^2$ dropped below the
given limit of $\chi^2$. We neglect here the so-called local optimizer such as
LM and SA algorithm because the efficiency of these algorithms depends strongly
on the starting values.}
  \label{tab:PerfTable}
\end{table}

In Table~\ref{tab:PerfTable} the total cost (in function evaluations) for each
test function (Rastrigin, Rosenbrock, and Himmelblau function) for each global
optimization algorithm is shown. As mentioned above, the total cost of an
algorithm, i.e., the number of function evaluations depends strongly on many
parameters. Therefore, Table~\ref{tab:PerfTable} can give only a very rough
overview of the efficiency for each algorithm. For example, the INS algorithm is
quite efficient in finding the global minimum of the Rastrigin function but less
efficient in finding the global minimum of the Rosenbrock function.

In general, the global optimization algorithms are more efficient in finding the
areas of global minima. But they are less efficient in finding the exact
properties of the global minima. For that purpose, local optimizers, such as the
LM or SA are much more efficient. The combination of different algorithms is the
best strategy to find the complete description of the experimental data.

    \subsection{Algorithm chain}\label{subsec:algchain}

\begin{table}[b]
  \caption{Best parameter set with the corresponding confidence intervals
($\chi^2 = 2.5576 \cdot 10^{-5}$) after applying an algorithm chain.}
  \centering
  \begin{tabular}{c|c|c|c|c}
    \hline
    parameter & value at &  error                 & error                 & log\\
    name      & minimum  &  left                  & right                 & (evidence)\\
    \hline
    $x_1$     & $0.9954$ &  $3.43 \cdot 10^{-2}$  & $2.85 \cdot 10^{-2}$  & $-3.309$\\
    $x_2$     & $0.9905$ &  $7.38 \cdot 10^{-2}$  & $8.21 \cdot 10^{-2}$  & $-2.846$\\
    \hline
    \hline
  \end{tabular}
  \tablefoot{We used an algorithm chain consisting of the bees, the SA and the
error estimation algorithm, to determine the global minimum of the Rosenbrock
function. The bees algorithm requires only a specification of the range for each
free parameter. Here, the parameters $x_1$ and $x_2$ vary between -5 and +5.}
  \label{tab:AlgChain}
\end{table}

Therefore, MAGIX includes the possibility to send the results of the
optimization process performed by one algorithm to another optimization loop
through some different algorithm. As mentioned above, the SA as well as the
LM algorithm require starting values of the parameters that are optimized,
i.e., the user has to find a good fit by hand before applying these algorithms
produces useful results. Often, the location of the minimum can be guessed with
sufficient accuracy to give good starting values, but sometimes one is
completely in the dark. Using an algorithm chain, the user can first apply one
of the swarm algorithms, e.g., the bees or NS algorithm, to determine the
starting values for the subsequent local optimization algorithm using SA or the
LM algorithm. As shown in Fig.~\ref{fig:AlgChain}, we used an algorithm chain to
determine the global minimum of the Rosenbrock function where we first applied
the bees algorithm to explore the landscape of the problem. Using the best
result of the bees algorithm (the parameter set that corresponds to the lowest
$\chi^2$ value, here $x_1=1.7044$ and $x_2=2.8679$) as starting point for the
SA algorithm, we find the global minimum of the Rosenbrock function. MAGIX does
not only allow one to use the best but also the second best etc. result of a
swarm algorithm as starting values for other algorithms. Therefore, we are
able to find multiple minima of models that behave as the Himmelblau function.

To determine the confidence intervals for the parameters that are optimized, the
user has to set the error estimation algorithm as the last algorithm of the
algorithm chain. The confidence intervals for the parameters that are used in
the example shown in Fig.~\ref{fig:AlgChain} are given in
Table~\ref{tab:AlgChain}. The upper right panel of Fig.~\ref{fig:AlgChain} shows
the probability for a minimum along the $x_1$ axis holding parameter $x_2$
fixed at $x_2 = 0.9905$. The lower right panel of Fig.~\ref{fig:AlgChain}
shows the probability for a minimum along the $x_2$ axis holding parameter $x_1$
fixed at $x_1 = 0.9954$. These two panels clearly show that there is only
one minimum within the given parameter range.

    \section{Examples}

   \begin{figure}[!b]
      \centering
       \includegraphics[width=0.47\textwidth]{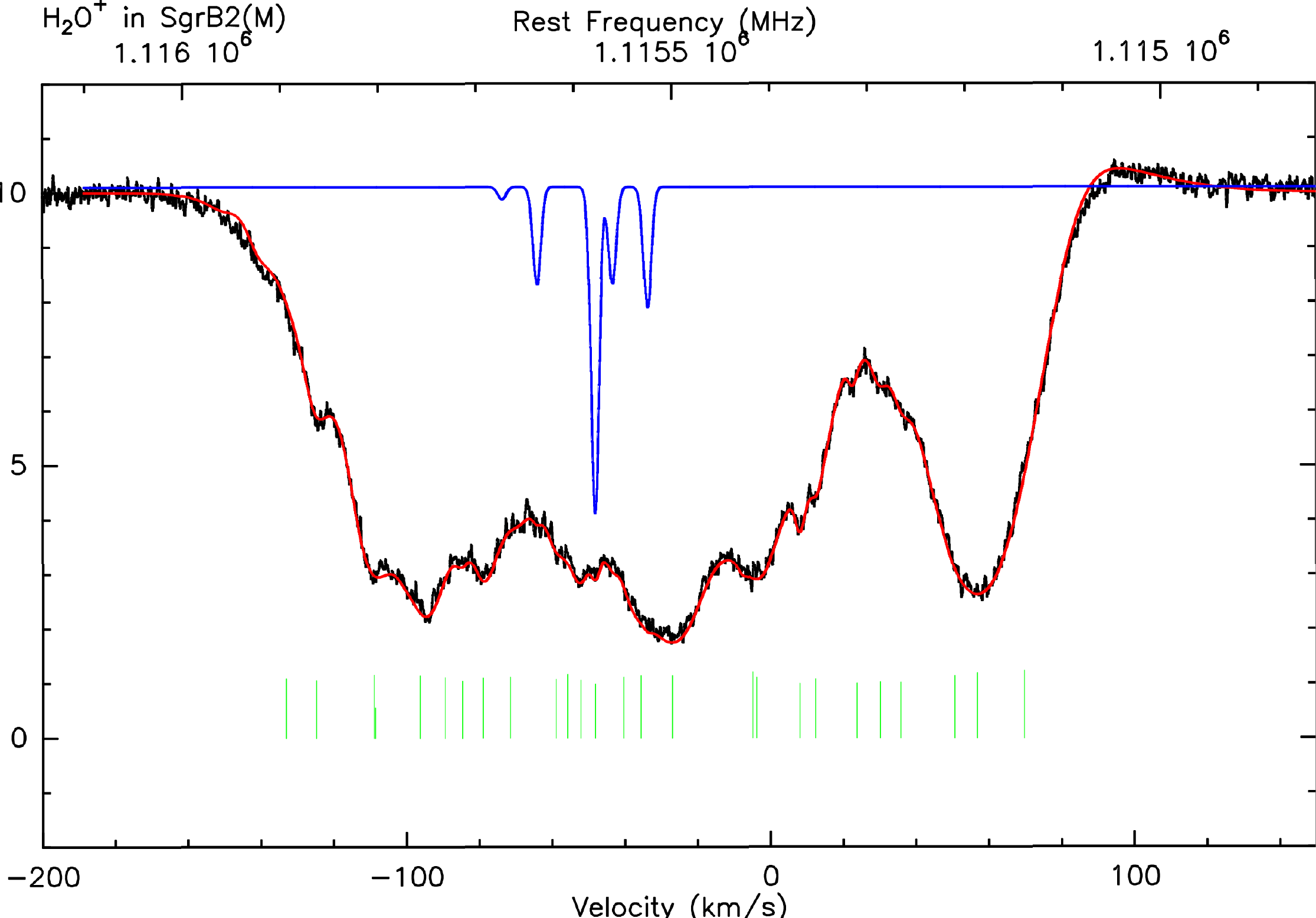}\\
      \caption{Fit of H$_2$O$^+$ using myXCLASS and MAGIX. In blue one specific
              component depicts the intrinsic hyperfine structure of the lines,
              while the green bars locate the positions of the velocity
              components.}
      \label{fig:Example}
    \end{figure}

The first astrophysical application of MAGIX was fitting Herschel/HIFI spectra
of an ortho-H$_2$O$^+$ line in Band 4b with myXCLASS\footnote{The myXCLASS
program (\url{https://www.astro.uni-koeln.de/projects/schilke/XCLASS}) accesses
the CDMS (\citet{Müller1,Müller2} \url{http://www.cdms.de}) and JPL
(\citet{Pickett1} \url{http://spec.jpl.nasa.gov}) molecular data bases. The
calculated spectrum is a solution of the radiative transfer equation in one
dimension (detection equation)
\begin{equation}
  T(\nu) = \sum_m \sum_c \eta \left( \theta_{m,c} \right) \left[ J \left(T_{ex}^{m,c} \right) - J \left(T_{bg} \right) \right] \left( 1 - e^{-\tau(\nu)^ {m,c}} \right),
\end{equation}
where the first sum goes over all molecules $m$ and the second sum runs over the
corresponding components $c$ of molecule $m$. $\tau(\nu)^{m,c}$ indicates the
optical depth of the component $m,c$ and $\eta \left( \theta_{m,c} \right)$
represents the corresponding beam-filling factor.}, in particular absorption
lines toward SgrB2(M) from the HEXOS~GT~KP \citet{Schilke1}. Here, the line
shape is determined by the hyperfine structure and the velocity structure,
resulting in a complicated pattern. The fit was performed using the LM
algorithm. As can be seen in Fig.~\ref{fig:Example}, the match is convincing,
but questions about the uniqueness of the fit and confidence intervals for the
free parameters, about 100 in this case do arise.

   \begin{figure}[t]
      \centering\includegraphics[width=0.45\textwidth]{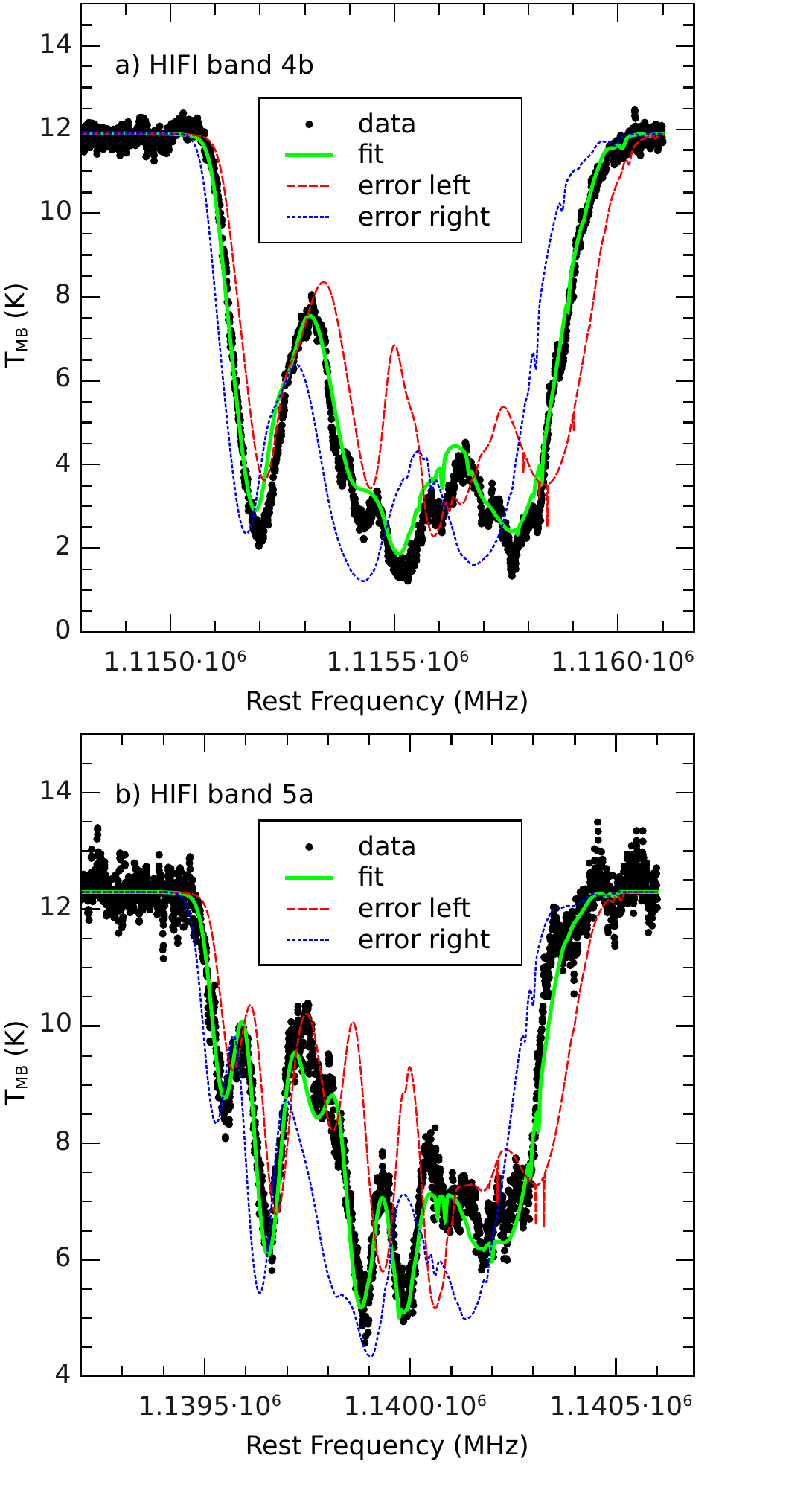}\\
      \caption{Fit of ortho H$_2$O$^+$ ground state lines in HIFI bands a)~4b
and b)~5a toward SgrB2(M) with myXCLASS using an algorithm chain consisting of
the genetic, the SA, and the error estimation algorithm. Owing to the high
number of free parameters (78) it is not possible to show the distribution of
the parameter values as in Figs.~\ref{fig:LM}~-~Fig.~\ref{fig:AlgChain}.}
      \label{fig:Sgrb2Fit}
    \end{figure}

Additionally, we fit HIFI bands 4b and 5a toward SgrB2(M) \citet{Schilke1} with
myXCLASS, simultaneously using an algorithm chain starting with the genetic
algorithm and 78 free parameters (we used 26 velocity components where each
component has three free parameters.). Here, we used the parameter values of the
best fit-result of the genetic algorithm as starting values for the SA. At the
end of the algorithm chain, we applied the error estimation algorithm to
determine the left and right error for each optimized parameter. The final
result is shown in Fig.~\ref{fig:Sgrb2Fit}, where the dashed red (blue) lines
indicate a model where we reduced (increase) the free parameter values of the
best fit by the left (right) error of each free parameter. Clearly, we
achieve an excellent description of the absorption lines and quantify the
uncertainty of the model. In contrast to the previous fit shown in
Fig.~\ref{fig:Example}, where only one data set was fitted, we defined no
starting values but only ranges for each free parameter and fitted two bands
simultaneously.

    \section{Conclusions}

The MAGIX package presented in this paper is a very helpful tool for modeling
physical and chemical data using an arbitrary external model program. It
represents a highly flexible toolbox where in principle any theoretical external
model program can be plugged in. Thus, MAGIX can be used to optimize the
description of even non-astrophysical data. Furthermore, MAGIX is able to
explore the landscape of the $\chi^2$ function without the knowledge of starting
values and can calculate probabilities for the occurrence of minima. Therefore,
MAGIX can find multiple minima and give information about confidence intervals
for the parameters. Additionally, the MAGIX package contains the possibility to
combine algorithms in a so-called algorithm chain and make use of the advantages
of the different algorithms included in the package. If the external model
program fulfills certain requirements, MAGIX can run in a parallel mode to speed
up the computation.

    \begin{acknowledgements}
          We acknowledge funding from Bundesministerium f\"{u}r Bildung und
Forschung (BMBF) through the ASTRONET Project CATS. We are grateful to
Fr\'ed\'eric~Boone for helpful discussions and Steven~N.~Shore for fruitful
contributions.\\
    \end{acknowledgements}

    
\end{document}